\documentclass[notoc,letterpaper]{JHEP3} 
\usepackage{amsmath}
\usepackage{epsfig}
\usepackage{amssymb,amsfonts}

\def\D{{\cal D}}

\title{Understanding Beth, the Particulate Mass Functional}
\author{Wayne R. Lundberg\\University of Dayton, Dayton OH 
45469-0210\\ \email{wayne.lundberg@mailaps.org}}

\abstract{
A geometric relationship between loop quantum gravity and 
partitioned (triangulated) string theory is discussed. 
Combinatorial analysis reveals that three spatial and three 
curvature dimensions, intrinsic to the partitioned string, 
are necessary to replicate Standard Model particles and interactions. 
This analysis has established that particulate mass is 
determined by a functional relationship involving these six 
extra dimensions.  The combinatorial analysis involves 
non-commutative 3D-matrix algebra which forms the mathematical 
underpinnings of Dirac notation.  The functional relationship (symbolized 
by Beth) requires exponential, Randall-Sundrum, scaling to compute mass. 
Through the proper interpretation of complex gravity a cyclic cosmological
model is developed.  This formulation of cyclic cosmology inherently 
involves observed dark energy.  Thus, a comprehensive theory is 
constructed from geometric fundamentals which models both massive, 
oscillating neutrinos and the current epoch of mini-inflation.
}

\preprint{...}

\begin{document}

\section{Introduction}
Physical insight, regardless of the technical area, often takes the form of
diagrams upon which appropriate mathematical formulations are applied.  The 
appropriateness of the mathematics is determined by its ability to 
produce calculable predictions which may be confirmed by 
experiment.  From Newton to Feynmann, many deep understandings of physics
have been developed and communicated using diagrams constructed from 
fundamental principles.  A great deal of promising mathematical 
exploration and tool-building has occurred recently in high energy physics 
and cosmology.  These efforts have yet to reveal the fundamental 
principles upon which to build a successful, comprehensive theory.

This paper takes the approach that particle theory must be constructed from 
objects which are fundamental (or minimal) in four-space, much as a construct
of three points is minimal in two-space.  Further, Standard Model particle 
properties are
expected to be identifiable with geometric properties intrinsic to the 
theoretical constructs representing quarks and leptons.  Gravitation, or 
particulate mass, must also be intrinsic to the theoretical construct, as 
well as separable from the other particle properties.  This 
requires that the theory provide a comprehensive explanation of cosmology 
and astrophysical objects, particularly black holes.  The resulting 
theoretical architecture would thus constitute a 'theory of everything'.  
The approach is not expected to yield a wholly new extension of 
mathematics, but rather insight into how (and which) mathematical tools are 
applicable and how to assemble them with reasonable expectation of 
determining that they are appropriate.

Extrapolating from the minimal object in two-space, one observes that a 
tetrahedral construction of four points is minimal in three-space - an object
that doesn't experience time in the usual sense.  A triangular construction, 
spinning about an axis, is minimal in four-space.  Such a construction is 
identified with a gravitational quantum, by noting its geometrical relation 
to the basis for loop Quantum Gravity (QG).  Such a simple construct would 
yield a theory which is clearly insufficient to explain the large variety of 
fundamental 
particles necessary to (at least) replicate the Standard Model.  A spinning
triangle can be quantized by considering orthogonal spin axes, one along 
an edge and one perpendicular to it.  Thus any spinning triangle could be 
resolved as a superposition, or mixing, of these two fundamental 
objects.  One could consider a construction of pairs of such objects, 
although that architecture also is insufficient to (at least) replicate the 
Standard Model.  

The correct combinatorial algebra was determined by associating the two 
fundamental 4D-geometric objects with Rishons \cite{Rishon}:
"T", spinning perpendicular to an 
edge, and "V", spinning along an edge.  This association exploits the Rishon 
theory's ability to replicate the Standard Model combinatorially.  
The geometry of the composite object is named tripartite, meaning "having 
three partitions".  The perimeter is considered string-like, however, in order
to intrinsically incorporate particulate mass into the theoretical 
architecture, the construct must have a calculable cross-sectional area.   
The construct is, more precisely, a 1-brane whose mass may be determined
through application of quantum gravity.

Careful selection of a symbol which can be used to annotate the functional
relationship between the intrinsic dimensions of the tripartite string 
and its computable mass was called for.  The Hebrew character "Beth" was 
chosen because (a) it implies a new, "constructive", line of mathematical 
reasoning; and (b) its meaning, "bountiful", implies that a large number of
fundamental particles originate through it.

The theoretical architecture and its relationship to
major areas of mathematical exploration is discussed beginning with the
applicable combinatorial algebra.  Much of the language describes diagrams 
which, of necessity, could not be included.  Diagrams of these ideas
were presented at conferences held by the Division of Particles and Fields 
of the American Physical Society in August 2000 and May 2002.  The 
associated transparencies were posted on the DPF conference websites.


\section{Beyond Dirac Notation}
Following Dirac, \cite{Dirac} it is conventional to express $\alpha$ in 
terms of generalized Pauli matrices, $\sigma_1, \sigma_2, \sigma_3$, 
and a set of anti-commuting matrices, $\rho_1, \rho_2, \rho_3$. The time 
component of Dirac 4-vectors shall be set aside, or separated, in order
to establish a non-commutative algebra applicable to Quantum Chromodynamics 
(QCD).  Examination of SU(3) by developing an applicable 
3D-matrix algebra representation allows one to define appropriate 
algebraic restrictions, equivalent to X SU(2) X U(1), which replicate QCD 
objects and interactions.  
Let 
$\sigma_1=
\left( \begin{array}{l}
010\hfill\cr 100\hfill\cr 001\hfill\cr
\endarray \endgroup \right)$
and 
$\rho_1=\left( \begin{array}{l}
001\hfill\cr 100\hfill\cr 010\hfill\cr
\endarray \endgroup \right)$.
Operating on a generalized 3x3 matrix gives 
$\alpha _1\left( \begin{array}{l}
abc\hfill\cr  def\hfill\cr  ghi\hfill\cr
\endarray \endgroup \right)
=\rho _1\sigma _1\left( \begin{array}{l}
abc\hfill\cr  def\hfill\cr  ghi\hfill\cr
\endarray \endgroup \right)
=\left( \begin{array}{l}
ghi\hfill\cr  def\hfill\cr  abc\hfill\cr
\endarray \endgroup \right)$, 
which has the transpose 
$\left( \begin{array}{l}
gda\hfill\cr heb\hfill\cr ifc\hfill\cr 
\endarray \endgroup \right)$.  
Recognizing this as a 90$^o$ rotation of the original matrix leads to the 
definition $R\equiv \alpha _1^T\equiv (\rho _1\sigma _1)^T$.  

\FIGURE[pos]{\epsfig{file=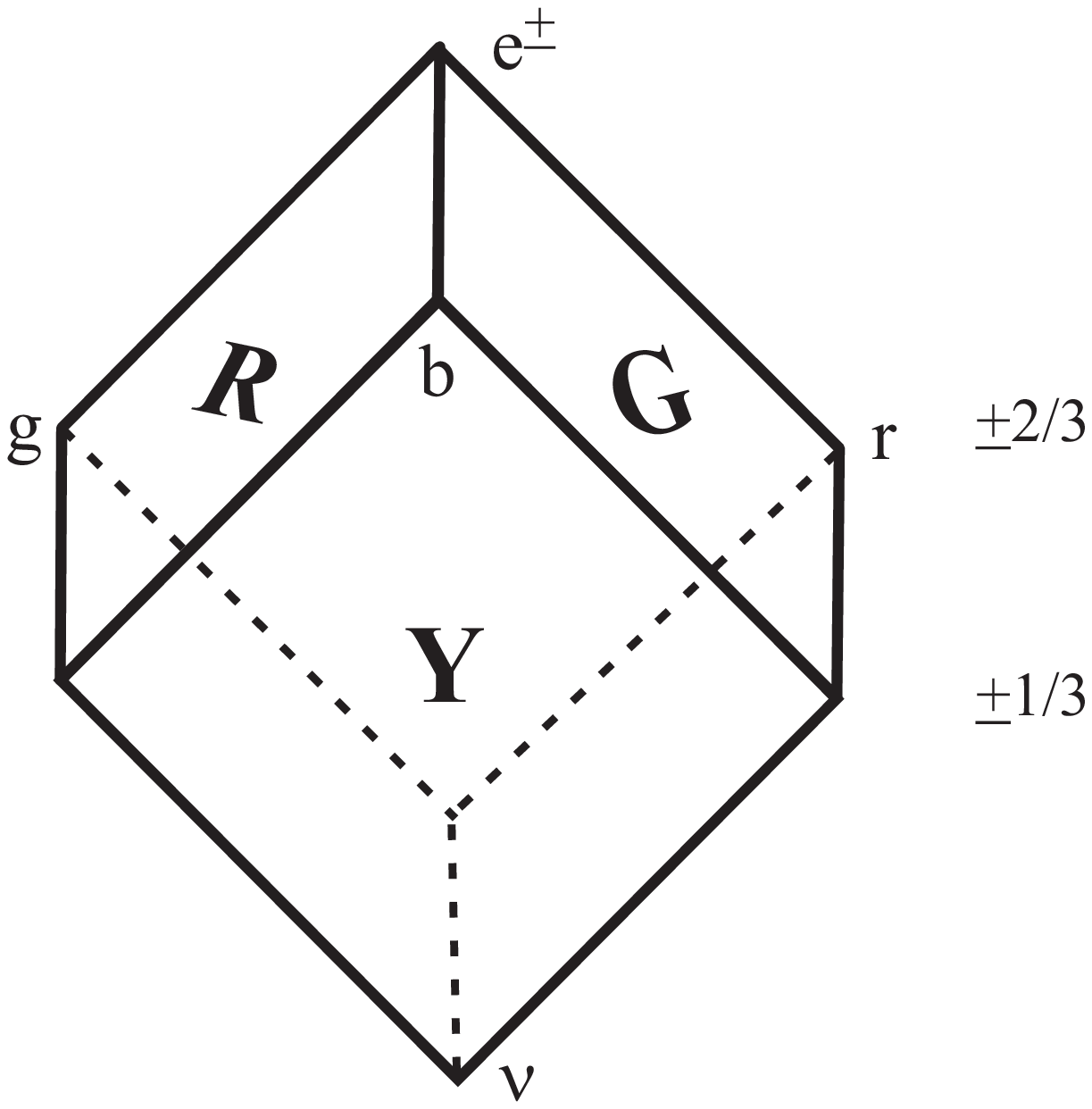,width=53mm}
 \caption{Charge and QCD symmetries applied to a 3x3x3 matrix.}}


The 3x3x3 matrix algebra notation is oriented on the cubical QCD charge 
and color symmetry diagram by assigning $R$ to the face opposite the red 
quark's vertex and $B, G$ to the faces in a clockwise fashion 
looking down from the top.  Anti-colors are simply labeled $M=$ magenta 
(anti-green), $C=$ cyan, $Y=$ yellow.  The usual rules for employing Dirac 
notation apply, except that complex notation is not required: antiparticles 
are modeled by preserving the twist of vertex elements.  

An $r\bar gb$ proton, ${\left| H_{r\bar gb} \right\rangle }$, is then 
explicitly modeled by 
${\left| {(G^-CGC^-\cdot C^-YCY^-\cdot Y^-GYG^-)^2} \right\rangle } $.
A path-integral approach was developed \cite{WRL1} which produces explicit 
formulae for modeling both strong and weak interactions.  The result was an 
algebra which is less generic than the equivalent string theoretic formula 
\cite{Witten}.  The operator
$G_{gb}=\left\langle {BR^-\left| {(CB^-C^-B)^2} \right| RB^- }
\right\rangle $, applied to ${\left| H_{r\bar gb} \right\rangle }$ 
from the $\it right$ produces an $rg\bar b$ proton.  
The 3D-matrix notation captures weak interactions just as explicitly as 
strong interactions, by noting that the 3D-matrix represents only one 
particle at a time.  In this way, the intermediate vector boson may be
modeled separately from the particles it interacts with.  
The general Dirac notation 
$ \left\langle {\alpha | r} \right\rangle 
\left\langle { r | \beta } \right\rangle $ becomes 
$ \left\langle {CR^-MG^-R^2 | (YR^-Y^-R)^2} \right\rangle
\left\langle { (RM^-R^-M)^2 | R^2GM^-RC^-} \right\rangle $ explicitly,
one of three similar 3D-matrix operators representing $W^+, W^-, Z^o$. 

The basic elements of 3D-matrix algebra are thus established, with 
symmetry-breaking restrictions employed to replicate SU(3) X SU(2) X U(1).  
These restrictions involve operators of the generic form: $(RM^-R^-M)^2$, 
which would permit a simpler notation to be defined.
The notation $\D_{3x3x3}$ is defined to mean any Standard Model quantum object
or interaction written in the foregoing Dirac-like algebra.  Although 
$\D_{3x3x3}$ serves to put 
QCD and QED on the same mathematical basis, it reveals no 'new physics'.  It
simply establishes the noncommutative algebra of quantum particulate states,
allowing a 'separable' commutative algebra to describe mass, time and energy.

\section{Extra Dimensions}
One can now demonstrate that $\D_{3x3x3}$ corresponds one-to-one with the 
fundamental, tripartite, modes of a string and its interactions.
An orthonormal orientation vector is defined such that the $R,G,B$ partitions
establish a particle's chirality.  In this regard it is more precisely a 
1-brane, closely related to a Typo IIA Orientifold \cite{Kaku}
Each string partition intrinsically has both a radial, $(r_R,r_G,r_B)$, and 
a curvature, $(\phi_R, \phi_G, \phi_B)$, dimension.  The color of a quark is
defined in accordance with the orientation of its $\it spin$ in relation to 
the colored partitions.  Since, by construction, there is two of one type 
'Rishon' and one of another, it is the odd one which establishes QCD color 
for the construct.  In this way a red up quark could be labeled VTT, by 
defining the positions in this variant of Rishon notation in RGB order.
One can now complete the correspondence to the tripartite string's intrinsic
geometry as discussed in the introduction.
The electric charge of quarks and leptons in the construct is easily 
replicated by identifying the "T" partition as having a 1/3 electric charge.
Identifying mass with the conformal area of the 1-brane completes the 
identification of particle properies as intrinsic properties of the construct.

Conformal area of the partitioned or triangulated string is naturally a 
functional relation of the extra spatial and curvature dimensions.
Exponential scaling of mass has been examined by Randall-Sundrum \cite{RS} 
theory and is incorporated in applicable QG models \cite{Gross,Rovelli}.  
Thus it is logical to consider the functional relation 
$\lambda \equiv \backslash Beth(r_R,r_G,r_B,\phi_R,\phi_G,\phi_B)$ when 
calculating particulate mass.
Further, the explicit formulation of $\D_{3x3x3}$ requires that standard 
calculations include a sum over all chromodymanic states of hadrons as 
well as standard tree-level diagrams to account for interactions.  This 
then leads to the relation:

\begin{equation}
m=\sum\limits_{trees} {\sum\limits_{states} {G_{\mu \nu }e^{-\lambda 
}|\left. {\D_{3x3x3}} \right\rangle }},
\end{equation}
which is valid for particles and condensed matter where 
$\sum |\left. {\D_{3x3x3}} \right\rangle \to 1$.  

\section{Quantum Gravity}

The method of calculating particulate mass using quantum gravity  
triangulation of a 2D surface is naturally applicable \cite{Rovelli}
to the three partitions of the string when it is regarded as the edge of 
a zero-thickness membrane surface.  
In this construct, energy is represented by string tension, which is 
altered when a photon is added.  Thus a photon must be an open (massless) 
string whose energy causes a change in string tension when it interacts
with any particle.  The photon can now be considered to mediate gravity as
follows: a change in string tension calls for equivalent, quantized change 
to partition radii $(r_R,r_G,r_B)$ and intrinsic curvature angles 
$(\phi_R, \phi_G, \phi_B)$.  The mass or energy of the interacting particle 
may be affected depending on whether the altered quanta combine to affect 
the conformal area of the closed 1-brane.  

A given particle
will have a unique 'shape', $(r_R,r_G,r_B,\phi_R,\phi_G,\phi_B)$, or 
low-energy string mode, as determined by its spin and substructure.  The 
intrinsic shape is analogous to determining the electron orbital shapes from 
QED and spherical harmonics.  Analysis to resolve this nonperturbative string 
shape exactly must be accomplished in light of the fundamental geometric 
construction as outlined above.
The case of weak interactions in which the identity, and mass, of the 
interacting particles is changed, requires that the mass functional Beth be
commutative.  Thus Beth is also dependant on the non-commutative algebra 
required to model QC/ED interactions.  An intermediating 
particle also acquires mass from its intrinsic geometry, although the geometry
is quite different in the case of a gluon.  

Electro-weak symmetry requires that massive bosons interact 
much like the photon in that they cause a quantum change in each interacting 
strings intrinsic curvature.  They also must change the sign of the intrinsic
curvature, $\phi$, in relation to spin axis, since a quantum change is 
required to change a particle's identity.  Dimensional analysis reveals that
weak interactions are mediated by bosons comprised of two tripartite string 
quarks in an intrinsically bound state.  Such a tightly bound state doesn't 
require another intermediating particle, so the compositeness is hidden.  

Strong interactions of the tripartite string would require that each 
quark change its spin axis in order to change color-identity.  The 
partitioned string model calls for such interacting quarks to stretch into 
a relatively long and thin 'dog-bone' configuration, since the 
dimensions of the quarks are much less than their interaction range.
Considering that the area of such a configuration is dominated by the
thin ribbon-like area attributed to the gluon supports the observations
that the proton's mass is mostly attributable to gluons.  The
architecture of this theory is directly comparable to the modified Lund 
theory, which uses an open string with thickness to model mass.

The QG-partitioned string has a fundamental combinatorial state which
serves to model the neutrino (VVV).  One should note that any closed loop 
has a torsion-free, preferred, coiled state which can be called a trecoil.
A trecoil is similar to a trefoil knot, except that a trefoil knot 
requires that the string be broken and re-attached.  In this construction, 
a trefoil knot represents a 'sterile' neutrino.  Considering the
entrained area of the trecoil string state to model its mass explains
why the neutrino's mass is so small.  Further, it is easy to construct
simple variations in which each third-coil is unfolded to create a
more massive, higher-generation, state of the neutrino.  In this way,
massive, oscillating neurinos are a natural part of the model.
\cite{Fukuda}

The construction of generalized string interaction vertices has been shown
to imply that strings have three partitions \cite{MKaku}.  Construction
of a four-string interaction involves a tetrahedron at the vertex - which may 
be interpreted as a Higg's Boson, given its role in mediating certain 
interactions.  Causality is preserved in this theory \cite{Seiberg}, since 
both mass and time are derived from a commutative operation.

The membrane surface is also a mathematical discontinuity in space-time:
when a particle passes across it, it actually experiences 
$R\leftrightarrow \alpha^{\prime} /R$.  The theory thereby includes a natural
model of antimatter, which is created during interactions when
a partition of an interacting matter string is 'pushed' across this 4-space 
discontinuity.  The existance and function of this discontinuity has 
important implications for both astrophysics and cosmology.  

\section{Astrophysics}
Standard astrophysical theory includes a mathematical singularity at the 
center of all black holes.  However, M/string theory clearly excludes such
singularities from any unified model.  Two distinct lines of reasoning 
say that a 'physical singularity', constructed as a Planck-scale 
tetrahedron, resolves the dilemma.  
(1) particles impinging on a black hole experience a 'slowing' of time, 
due to the extreme curvature of space-time.  A central singularity or 
discontinuity exists at which time 'stops' altogether.  A tetrahedral 
construction is 
deduced such that the singularity within a black hole is a minimal object 
in three-space.  Note that a tripartite string is not assumed, but necessary 
as a facet of the central tetrahedral discontinuity.  (2) Any standard model 
of a black hole (with a mathematical singularity) is bi-laterally symmetric, 
whereas one
with a tetrahedral kernel is not.  Matter which is relatively stopped in time 
has an observable motion relative to the black hole and thus our space-time.  
The Galactic Annihilation Fountain \cite{Dermer} and other artifacts are 
clearly bi-laterally asymmetric, and so {\it qualitatively} support the 
construct of a tetrahedral kernel over theories including a mathematical 
singularity.
Maldacena's conjecture should be developed, beyond proving 
consistency with R-S scaling \cite{Duff}, to understand the  
mathematics of AdS/CFT at a tetrahedral 'interface'. 

\section{Cosmology}

\FIGURE[pos]{\epsfig{file=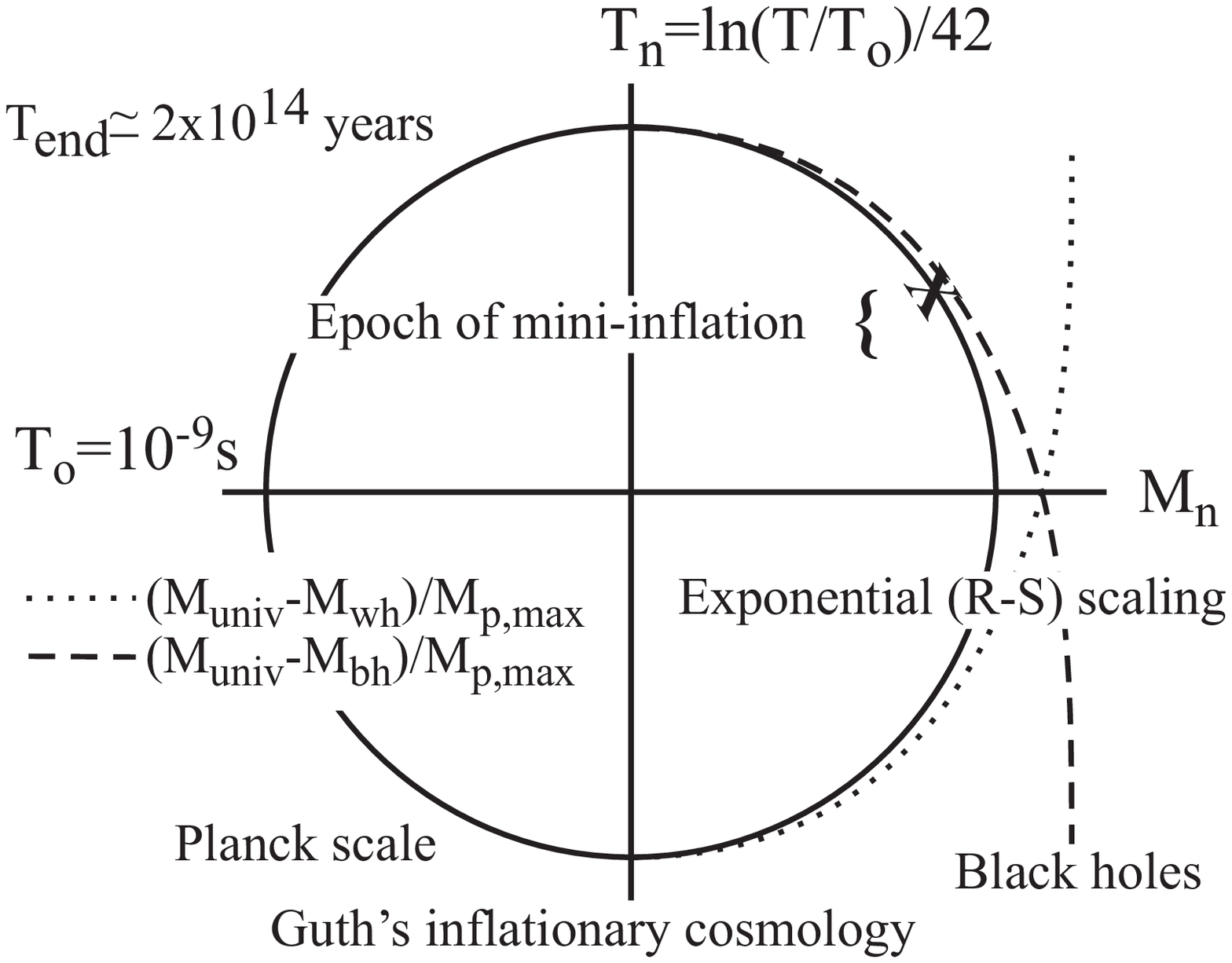,width=75mm}
  \caption{Cyclic cosmological diagram showing the relationship to 
crucial theoretical structures and recent observations.}}


An exponential (R-S) mass scale yields a cyclic cosmological model through 
consideration of complex gravity \cite{AC}, which is inherent to string 
theory.  String theory tells us that the usual gravitational tensor has an 
anti-symmetric component which is 
important over cosmological time-scales.  Including the anti-symmetric 
tensor in equation (3.1) yields: 
$\sum\nolimits_U {(G_{\mu \nu }+iB_{\mu \nu })e^{-\lambda }.}$

We now recognize that applying Euler's relation to this formula yields an
equation having the form
$\Omega _m+\Omega _\Lambda =1;$ where $\Omega_m$ is again area-like and 
both terms are scaled exponentially as in R-S theory.  
This approach yields a cyclic cosmology that incorporates the string 
theoretic result $R\leftrightarrow \alpha^{\prime} /R$ \cite{WRL2}.  This
important relation helps explain how the initial state of the universe came 
about.  In an 'inverted' space-time, the vast scale of the cosmos equates 
a sub-Planck scale 
state 'preceding' a 'spherically-inverted' big bang.
This theory also inherently includes the observed \cite{Perl} cosmological 
'constant' as an extremely slowly-varying parameter \cite{Shapiro}.
The theoretical requirement for an imaginary time-like metric, 
$e^{i\theta_{m,t}-\lambda}$, 
permits cosmological boundary conditions to be established which select 
the observed set of physical laws.  
This formulation also circumvents the coincidence problem through a 
simple mechanism relating $\Omega _m$ to $\Omega _\Lambda$ at all times 
\cite{Dalal}.

\section{Observation and Experiment}

This work qualitatively explains many recent observations 
\cite{Fukuda, Perl, Dermer} in a comprehensive
fashion that is consistent with developing modern theories.  Although the 
standard model may be able to be modified or interpreted to explain these 
observations separately, the unified approach given here has the added 
benefit of being based on geometric fundamentals.
Experimentation with high-energy electron-positron collisions should shed 
some light on the theorized shape of an electron.  Ongoing research and 
observation programs in the areas of oscillating neutrinos, galactic 
annihilation radiation and cosmology are likely to support this 
comprehensive theoretical approach.  

\section{Conclusion}

The formula $e^{i\theta_{m,t}-\lambda}|\left. {\D_{3x3x3}} \right\rangle$
provides the basis for a comprehensive theory of particle physics and 
cosmology.  Further work to refine the theory and calculate its effects 
should be within the realm of modern mathematics.


\begin{thebibliography}{99}

\bibitem{Rishon} H. Harari, "{\it Scales of Compositeness for Quarks
and Leptons}", {\it Festschrift in Honor of Yuval Ne'eman}, 113, 1982.

\bibitem{Dirac} G. Esposito, {\it Dirac Operators and Spectral
Geometry} (Cambridge Univ. Press, Cambridge, 1998).

\bibitem{WRL1} W.R. Lundberg, "{\it Topological Combinatorics of a 
Quantized String Gravitational Metric}", Proc. 7th APS/DPF, (1992) 1589, 
physics/9712042; Proc. Beyond the Standard Model III, (1992) 515.  

\bibitem{Witten} E. Witten {\it et al}, ed., {\it Quantum 
Fields and Strings: A Course for Mathematicians},
{Vol. 1} (American Mathematical Society and Institute for Advanced 
Study, Providence RI, 1999), 505.

%
\bibitem{Kaku} Z. Kakushadze, "{\it Orientiworld}", J. High Energy Phys. 
10 (2001) 031, hep-th/0109054 .

\bibitem{RS} L. Randall and R. Sundrum, "{\it An Alternative to
Compactification}", Phys. Rev. Lett. (1999) Vol. 83, 4690, hep-th/9906064.

\bibitem{Gross} D.J. Gross and A.A. Migdal, "{\it Nonperturbative
two-dimensional quantum gravity}", Phys. Rev. Lett. (1990) 
Vol. 64, 127.

\bibitem{Rovelli} L. Crane, A. Perez and C. Rovelli, "{\it Perturbative
Finiteness in Spin-Foam Quantum Gravity}", Phys. Rev. Lett. (2001), Vol. 87,
181301.

%
\bibitem{Fukuda} Y. Fukuda et al, "{\it Evidence for oscillation of 
atmospheric neutrinos}", Phys. Rev. Lett. (1998) Vol. 81, 1562, 
hep-ex/9807003 ({\it et seq}.).

\bibitem{MKaku} M. Kaku, "{\it Nonpolynomial closed-string field theory}",
Phys. Rev. D, Vol. 41, 3734, 1990.

\bibitem{Seiberg} N. Seiberg, L. Susskind and N. Toumbas, "{\it Space/time
non-commutivity and causality}", J. High Energy Phys. 06 (2000) 044.

\bibitem{Dermer} C.D. Dermer and J.G. Skibo, "{\it Annihilation Fountain 
in the Galactic Center Region}", Astroph. J. Lett., 
astro-ph/9705070 ({\it et seq}.).

\bibitem{Duff} M.J. Duff and J.T. Liu, "{\it Complementarity of the Maldacena
and Randall-Sundrum Pictures}", Phys. Rev. Lett. Vol. 85, 2052 (2000), 
hep-th/0003237.

\bibitem{AC} Ali Chamseddine, "{\it Complex Gravity and Noncommutative
Geometry}", Proc. 2000 Int'l Superstrings Conf., (2000) 119.

\bibitem{WRL2} W.R. Lundberg, "{\it A Cyclic Universe without Missing Mass:
Implications of $R\leftrightarrow \alpha^{\prime} /R$}", 
Proc. 9th APS/DPF, (1996) 1400, astro-ph/0007100.

\bibitem{Perl} S. Perlmutter et al, "{\it Measurements of Omega and Lambda
from 42 High-Redshift Supernovae}", Astrophys. J., astr-ph/9812133 
({\it et seq}.).

\bibitem{Shapiro} I.L. Shapiro and J. Sola, "{\it The scaling evolution of
the cosmological constant}", J. High Energy Phys. 02 (2002) 006.

\bibitem{Dalal} N. Dalal, K. Abazajian, E. Jenkins and A.V. Manohar,
"{\it Testing the Cosmic Coincidence Problem and the Nature of Dark Energy}",
Phys. Rev. Lett. Vol. 87 (2001) 141302, astro-ph/0105317. 

\end{thebibliography}
\end{document}